\documentclass[twocolumn,pre,superscriptaddress,amsmath]{revtex4}

\usepackage{graphicx}
\usepackage{dcolumn}   
\usepackage{bm}        

\newcommand{\eref}[1]{(\ref{#1})}
\newcommand{\figref}[1]{Fig.~\ref{#1}}
\newcommand{\tabref}[1]{Table~\ref{#1}}

\newcommand\ave[1]{\left\langle{#1}\right\rangle}
\newcommand\avek{\ave{k}}
\newcommand\avekn[1]{\ave{k^{#1}}}
\newcommand\aves[2]{{\left\langle{#2}\right\rangle_{\scriptstyle{#1}}}}
\newcommand\kmax{k_{(\text{max)}}}
\newcommand\sumone{\sum_{k=1}^\infty}
\newcommand\sumtwo{\sum_{k_1,k_2=1}^\infty}
\newcommand\sumthree{\sum_{k_1,k_2,k_3=1}^\infty}
\newcommand\sumfour{\sum_{k_1,k_2,k_3,k_4=1}^\infty}
\newcommand\rann{\medskip%
  \centerline{\sf --- Random-network approximation ---}%
  \smallskip}  
\newcommand\rannn[1]{\medskip%
  \centerline{\sf --- Random-network approximation {#1} ---}%
  \smallskip}  

\begin{document}

\title{Large-scale structure of a nation-wide production network}

\author{Yoshi Fujiwara}
\affiliation{NiCT/ATR CIS, Applied Network Science Lab, Kyoto
  619-0288, Japan}
\author{Hideaki Aoyama}
\affiliation{
  Department of Physics, Kyoto University, Kyoto 606-8501, Japan}


\begin{abstract}
  Production in an economy is a set of firms' activities as suppliers and
  customers; a firm buys goods from other firms, puts value added and
  sells products to others in a giant network of production. Empirical
  study is lacking despite the fact that the structure of the
  production network is important to understand and make models for
  many aspects of dynamics in economy. We study a nation-wide
  production network comprising a million firms and millions of
  supplier-customer links by using recent statistical methods
  developed in physics. We show in the empirical analysis scale-free
  degree distribution, disassortativity, correlation of degree to
  firm-size, and community structure having sectoral and regional
  modules. Since suppliers usually provide credit to their customers,
  who supply it to theirs in turn, each link is actually a
  creditor-debtor relationship. We also study chains of failures or
  bankruptcies that take place along those links in the network, and
  corresponding avalanche-size distribution.
\end{abstract}


\maketitle

\section{Introduction}\label{sec:intro}

The physics community has recently witnessed considerable development
of statistical methods for quantifying large networks, including
biology, information, technology, economics and society
\cite{albert2002smc,dorogovtsev2002en,newman2003sfc}. The development
enables one to quantify statistical features, modular and
heterogeneous structures of large networks that are not amenable even
to visualization.

{\it Production network\/} in economics refers to a line of economic
activities in which firms buy intermediate goods from ``upstream''
firms, put value added on them, and sell the goods to ``downstream''
firms. Net sum of value added in the whole network is basically the
net total production in a nation, that is, gross domestic product
(GDP).

Consider a ship manufacturer, for example. The manufacturer buys a
number of intermediate goods including steel materials, mechanical and
electronic devices, etc.~and produces ships. The firm puts
value added on the products in each process of production. In the
upstream side of the ship manufacturer, a processed steel
manufacturer could be present, which in turn buys intermediate goods such as raw steel
and fabricating machines. The steel manufacturer may supply its
products to car manufactures as well. Similarly in the downstream
side, including retail and wholesale firms.

The entire line of these processes of putting value added in turn,
therefore, forms a giant web of production ranging from upstream to
downstream, ultimately down to consumers. Each firm needs labor and
financing in addition to intermediate goods, and utilizes these inputs
to produce outputs in {\it anticipation\/} for return of profits.
Thus, a real economy has its driving force in production, and is fueled by
labor and financing.

There are studies based on theoretical models of production networks, notably in
overlapping communities between economists and physicists. They include
inventory dynamics \cite{bak1993soc,scheinkman1994soc,nirei2000sid}
(see also \cite{bak1996hnw}), suppliers/customers dynamics
\cite{weisbuch2007pnf}, and credit-chain model
\cite{battiston2007ccb,lorenz2008srn,sieczka2009cfb}.
These promising works are, unfortunately, not based on
any empirical study on the structure of production networks.
While economic networks of inter-bank 
\cite{demasi2006fmi,iori2008nai,kyriakopoulos2009nea}, banks-firms 
\cite{demasi2009ajc} financing relations, and inter-firm ownerships
\cite{garlaschelli2005sew,glattfelder2009bcn} are studied based on
recent empirical data (see \cite{schweitzer2009en} and references
therein), there exists a gap to be filled in the study of inter-firm
credit and supplier-customer relations. So it is highly desirable to
investigate the structure of production network on a nation-wide scale
to develop insightful models, but such a study has been considered a
formidable task so far.

The present paper precisely performs such an empirical study of the
large-scale structure of a production network that comprises most
firms in a nation as nodes, and supplier-customer relations as links.
We will find that the network is not regular nor random, but possesses
scale-free degree distribution, disassortativity, as well as other
statistical properties, and structural modules that depend on
industrial sectors and geographical regions with highly varying
modular sizes.

The structural heterogeneity would have many important consequences in
{\it dynamics\/} on a production network. For example, demand by firms
and consumers downstream will propagate upstream; when foreign
countries increase demand for ships, it will result in a growing output
of ship manufacturers, which possibly stimulate production of processed
steel, raw metals, related machines etc.~in upstream firms. This
propagation will not take place homogeneously but heterogeneously.

Conversely, decreasing demand also causes a chain reaction. An
individual firm's profit is equal to sales {\it minus\/} costs. 
It may use factors of production in anticipation of profit, but
always faces uncertainty in ex-post demand, labor and financial costs,
price change of intermediate goods, and so forth.
Only {\it a posteriori\/}, profits are determined through the
interaction of firms in the production network. Once a firm goes into
a state of financial insolvency or bankruptcy, its upstream firms have its
balance-sheets deteriorated by losing fractional profits, and may
eventually go into bankruptcy. We will show that such a chain of
failure is by no means negligible, due to the network structure,
and so that this has a considerable effect at
macroeconomic activity.

In Section~\ref{sec:data}, we describe how nodes and links are
surveyed and recorded in our dataset of a production network,
in addition to another dataset of an exhaustive list
of bankruptcies occurred in the network over a certain period of time. In
Section~\ref{sec:stat}, we study the structure of the production
network by employing statistical methods in
\cite{albert2002smc,dorogovtsev2002en,newman2003sfc}. In
Section~\ref{sec:commun}, we extract community structure in the
network. In Section~\ref{sec:bankrupt}, we examine the chains
of bankruptcies with a focus on avalanche-size distribution.
After a brief discussion in Section~\ref{sec:discussion},
we conclude in Section~\ref{sec:summary}.

\section{Supplier-customer links and bankruptcy data}\label{sec:data}

Let us say that a directional link is present as $A\rightarrow B$ in a
production network, where firm $A$ is a supplier to another firm $B$,
or equivalently, $B$ is a customer of $A$. While it is difficult to
record every transaction of supply and purchase among firms, it is
also pointless to have a record that a firm buys a pencil from
another. Necessary for our study are data of links such that the
relation $A\rightarrow B$ is crucial for the activity of one or both
$A$ and $B$. If at least one of the firms at either end of a link
nominates the other firm as most important suppliers or customers, then
the link should be listed.

Our dataset for supplier-customer links is based on this idea. Tokyo
Shoko Research, Inc., one of the leading credit research agencies in
Japan, regularly gathers credit information on most of active firms
through investigation of financial statements and corporate documents,
and by hearing-based survey at branch offices located across the
nation. Financial and credit information of individual firms are
compiled in commercially available databases. The credit
information of individual firm includes its suppliers and customers, up to
24 companies for each, that are considered to be most crucial for its
business activities.
We assume that the links playing important roles in the
production network are recorded at either end of each link as we
describe above, while we should understand that it is possible to drop
relatively unimportant links from the data. Although amounts of
transactions provide information of weights on links, that is of relative
importance regarding suppliers and customers, such data are only partially
available at the moment. We simply ignore the weights in this paper.
It is also remarked that the financial sector is under-represented in
the database as those financial companies' links are not included.

We have a snapshot of production networks compiled in September
2006. In the data, the number of firms is roughly a
million, and the number of directional links is more than four
million (see Section~\ref{sec:stat}). The set of nodes in the network
covers essentially most of the domestic firms that are active in the
sense that their credit information is required. Attached to each
firm is financial information of firm-size, which is measured as
sales, profit, number of employees and their growth, major and minor
classification into industrial sectors, details of products, the firm's
banks, principal shareholders, and miscellaneous information
including geographical location.
In particular, the industrial sectors are classified into more than
1,200 industries and are categorized hierarchically into 19 divisions,
97 major groups and more than 400 minor groups (see \tabref{tab:sect}).  For
example, the manufacturing sector (F) is classified into 24 major
groups as tabulated in \tabref{tab:sect_major}. Each firm has industry
classification according to the sector it belongs to as primary (also
secondary and tertiary, if any) industry.

\begin{table}
  \caption{\label{tab:sect}%
    Classification of industrial sectors.
    Third column shows numbers of major-groups/groups/industries
    classified in each division. Fourth column are fractional numbers
    of firms in the divisions according to primary industry of each
    firm in the dataset of September 2006.}
\begin{ruledtabular}
\begin{tabular}{lllr}
  & divisions\footnotemark[1] & \#class.\footnotemark[1] & \#firms(\%) \\
  \hline
  A & agriculture & 1/4/20 & 0.53 \\
  B & forestry & 1/5/9 & 0.05 \\
  C & fisheries & 2/4/17 & 0.11 \\
  D & mining & 1/6/30 & 0.18 \\
  E & construction & 3/20/49 & 29.92 \\
  F & manufacturing & 24/150/563 & 17.69 \\
  G & electricity/gas/heat/water & 4/6/12 & 0.06 \\
  H & information/communications & 5/15/29 & 2.42 \\
  I & transport & 7/24/46 & 3.54 \\
  J & wholesale/retail trade & 12/44/150 & 29.07 \\
  K & finance/insurance & 7/19/68 & 0.65 \\
  L & real estate & 2/6/10 & 2.61 \\
  M & food establishments & 3/12/18 & 1.46 \\
  N & medical/health care/welfare & 3/15/37 & 1.03 \\
  O & education/learning support & 2/12/33 & 0.36 \\
  P & compound services\footnotemark[2] & 2/4/8 & 0.64 \\
  Q & services\footnotemark[3] & 15/68/164 & 9.45 \\
  R & government\footnotemark[3] & 2/5/5 & 0.18 \\
  S & unable to classify & 1/1/1 & 0.03 \\
  \hline
  & Total & 97/420/1,269 & 99.98 \\
\end{tabular}
\end{ruledtabular}
\footnotetext[1]{%
  Japan Standard Industrial Classification, Rev.~11, March 2002:
  {\tt http://www.stat.go.jp/english/index/seido/sangyo/index.htm}}
\footnotetext[2]{%
  Government-affiliated postal services, and agriculture, forestry,
  fisheries and business cooperative associations.}
\footnotetext[3]{%
  Not elsewhere classified.}
\end{table}

\begin{table}[tbhp]
  \caption{\label{tab:sect_major}%
    24 major groups of the manufacturing sector (F).}
\begin{ruledtabular}
\begin{tabular}{clr}
  id & major group & \#firms(\%) \\
  \hline
  09 & foods & 10.15 \\
  10 & beverages, tobacco and feed & 1.95 \\
  11 & textile mill products, except apparel/related & 3.08 \\
  12 & apparel and related finished products & 5.00 \\
  13 & lumber and wood products, except furniture & 3.57 \\
  14 & furniture and fixtures & 2.44 \\
  15 & pulp, paper and paper products & 2.90 \\
  16 & printing and allied industries & 6.51 \\
  17 & chemical and allied products & 3.28 \\
  18 & petroleum and coal products & 0.29 \\
  19 & plastic products, except otherwise classified & 4.98 \\
  20 & rubber products & 1.10 \\
  21 & leather tanning, leather/fur products & 0.83 \\
  22 & ceramic, stone and clay products & 5.09 \\
  23 & iron and steel & 1.97 \\
  24 & non-ferrous metals and products & 1.35 \\
  25 & fabricated metal products & 12.30 \\
  26 & general machinery & 13.83 \\
  27 & electrical machinery, equipment and supplies & 5.08 \\
  28 & information and communication electronics & 1.33 \\
  29 & electronic parts and devices & 2.57 \\
  30 & transportation equipment & 3.19 \\
  31 & precision instruments and machinery & 2.06 \\
  32 & miscellaneous manufacturing industries & 5.14
\end{tabular}
\end{ruledtabular}
\end{table}

In addition, we use a database that records ``dead'' firms, namely
business failure or bankruptcy. This dataset is an exhaustive list of
bankrupted firms since October 2006 for one year, corresponding to
the snapshot of the network. The data is exhaustive in the sense that any
bankrupted firm with a total amount of debt exceeding 10 million yen
(roughly 70 thousand euro or 100 thousand US dollar)
is listed therein. Each record includes the
date of failure, the total amount of debt when bankrupted and categorized
causes of bankruptcy. The dataset has high quality and its statistical
tabulation is employed by the Statistics Bureau of government for an
official statistics. In the production network, 0.5\% to 1\% of nodes
exit in a year due to failure (see Section~\ref{sec:bankrupt}). Thus,
by combining the two datasets of supplier-customer links and actual
failures, one has an opportunity to do an empirical study of the
dynamics of failure on the production network. And this point differs
from the previous studies on the Japanese production network including
\cite{saito2007dfl,ohnishi2009haj}.

\section{Structure of production network}\label{sec:stat}

\subsection{Global connectivity}\label{sec:connectivity}

The production network as a directed graph is not
uni-directional from upstream to downstream, but is highly
entangled depending on the products and services that each firm
produces. Let us first examine the global connectivity by using a
similar graph-theoretical method as was performed in the study of the
hyperlink structure of the world-wide web \cite{broder2000gsw}.
The following numbers are for the dataset of September 2006, which
contains 1,019,854 firms as nodes of the network excluding all
the bankrupted firms before the month.

From a directed graph, one can obtain an undirected graph by simply
ignoring the direction of links. A weakly connected component of
the directed graph refers to a connected component in the
undirected counterpart.
The production network has a giant weakly connected component
(GWCC) comprised of 99.0\% (1,009,597 nodes) of the whole. The rest
are disconnected components, all of which are smaller than 12 in size.

A strongly connected component (SCC) in a directed graph is a set of
nodes such that for any pair of nodes $u$ and $v$ in the set there is
a directed path from $u$ to $v$. There exists a giant SCC having the
size of 45.8\% of the GWCC (462,563 nodes). Calling it GSCC,
the GWCC turns out to be decomposed into mutually disjoint parts as
$\text{GWCC}=\text{GSCC}+\text{IN}+\text{OUT}+\text{TE}$, where
$\text{IN}$ is the set of non-GSCC nodes, from which one can reach
a node (so all the nodes) in the GSCC. Symmetrically, $\text{OUT}$ is
the set of non-GSCC nodes, to which one can go from any node in the
SCC. And TE is the rest of the GWCC, called tendrils, which consists
of nodes that have no access to the GSCC and are not reachable
from it. Hanging off IN and OUT are tendrils containing nodes that are
reachable from portions of IN, or that can reach to portions of OUT,
without passing through the SCC. See Figure~6 in
\cite{dorogovtsev2002en} understanding their definitions for giant 
$\text{GIN}$ and $\text{GOUT}$ as
$\text{GIN}=\text{IN}+\text{GSCC}$ and
$\text{GOUT}=\text{OUT}+\text{GSCC}$.
The IN, OUT and TE are composed of 18.0\% (182,018), 32.1\% (324,569)
and 4.0\% (40,447) nodes, respectively, in the GWCC.

By comparing the abundance of industrial divisions in each of these
giant components, we observe that in the portion of IN the numbers of
firms in the sectors of real estate~(L), forestry~(B),
information and communications~(H) are larger
when compared with the corresponding sectors in the SCC. In the portion of OUT
more abundant are medical, health care and welfare~(N), food
establishments~(M), education~(O). This fact is reasonable, because
these industries are located either in the upstream or in the
downstream. Nevertheless, all industries are basically embedded in the
SCC with entanglement. We shall study community structure in
Section~\ref{sec:commun}.

The diameter of a graph is the maximum length for all ordered pairs
$(u,v)$ of the shortest path from $u$ to $v$. The average distance is
the average length for all those pairs $(u,v)$. We found that
the average distance is 4.59 while the diameter is 22.

\subsection{Degree distribution}\label{sec:degree}

In the rest of this paper, we focus on the GWCC ignoring small
disconnected components. Denoting the numbers of nodes and links by
$N$ and $M$ respectively, they are for the GWCC
\begin{align}
  N &= 1,009,597\ , \label{num_N}\\
  M &= 4,041,442\ . \label{num_M}
\end{align}

\begin{figure}[htbp]
  \includegraphics[width=0.40\textwidth]{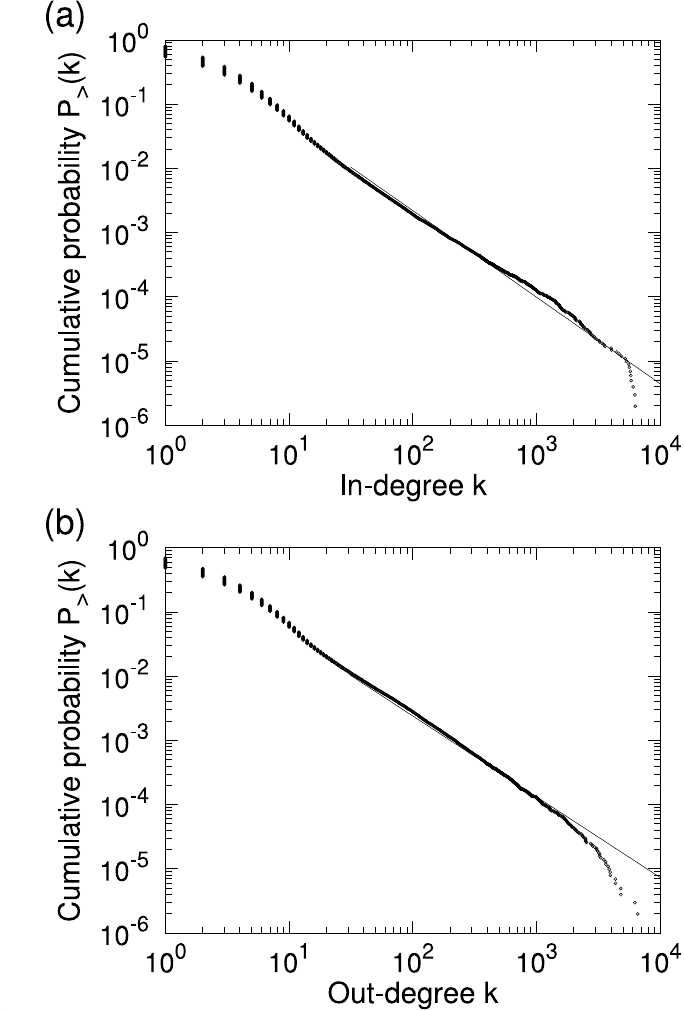}
  \caption{%
    (a)~Cumulative distribution for in-degree (number of suppliers).
     A power-law distribution $P_>(k)\propto k^{-\mu}$ is fitted
     by MLE with $\mu=1.35\pm0.02$ (a solid line).
    (b)~Same for out-degree (number of customers). The line is for
    $\mu=1.26\pm0.02$.%
  }
  \label{fig:deg_cpdf}
\end{figure}

A firm has suppliers for and customers of it, whose numbers are
in-degree and out-degree, respectively, according to our definition of
link direction. We show that both have a long-tail distribution.
Denoting the degree distribution by $P(k)$, cumulative distribution is
written as $P_>(k)=\sum_{k'=k}^\infty P(k')$. We plot the cumulative
distributions for in and out-degrees in \figref{fig:deg_cpdf}.

Both for in-degree and out-degrees of a firm, the distribution has a
heavy tail that can be characterized by a power-law $P_>(k)\propto k^{-\mu}$.
We estimated the exponent $\mu$ by maximum likelihood (MLE), {\it i.e.\/}
the Hill's estimate \cite{hill1975sga}, in a tail-region $k>k_*$. In
\figref{fig:deg_cpdf}, the estimates are shown for $k_*=40$, namely
$\mu=1.35\pm0.02$ for in-degree and $\mu=1.26\pm0.02$ for out-degree,
by solid lines. Here the errors correspond to $1.96\sigma$ (99\%
significance level) of the estimated standard errors $\sigma$.

The first two moments of in/out-degree are
\begin{align}
  & \ave{k_{\text{in}}}\equiv\ave{k_{\text{out}}}=4.003\ , \\
  & \ave{k_{\text{in}}^2}=1.041\times10^3\ ,\quad
  \ave{k_{\text{out}}^2}=1.036\times10^3\ .
\end{align}
For the undirected graph, we have
\begin{align}
  \avek &= 2M/N=8.006\ ,  \label{k1}\\
  \avekn2 &= 3.070\times10^3\ .  \label{k2}
\end{align}

Firms with largest in-degrees belong to the sectors of manufacturing
and construction among others, including heavy industry, electrical
machinery (e.g.~Hitachi, Mitsubishi, Panasonic, Toshiba), automobiles
(Toyota, Nissan, Honda), metal production, and so on. Large construction
companies are also included. Firms with the largest out-degrees are
worldwide traders, distributors of construction-related materials,
metals, petroleum, mechanical and electrical instruments, and general
wholesale companies, as well as the manufacturing firms mentioned for
in-degrees.

\subsection{Correlation to firm-size}\label{sec:firm_size}

The number of suppliers/customers of a firm depends obviously on
firm-size, an important attribute. A large firm likely possesses numerous
suppliers to buy various intermediate goods; similarly it has a number
of customers to sell its products to so that it has the large size.
Firm-size can be measured in different ways, basically by {\it
  stock\/} variables (total-asset, number of employees, etc.) or {\it
  flow\/} variables (sales, profits, etc.).

\begin{figure}[htbp]
  \includegraphics[width=0.39\textwidth]{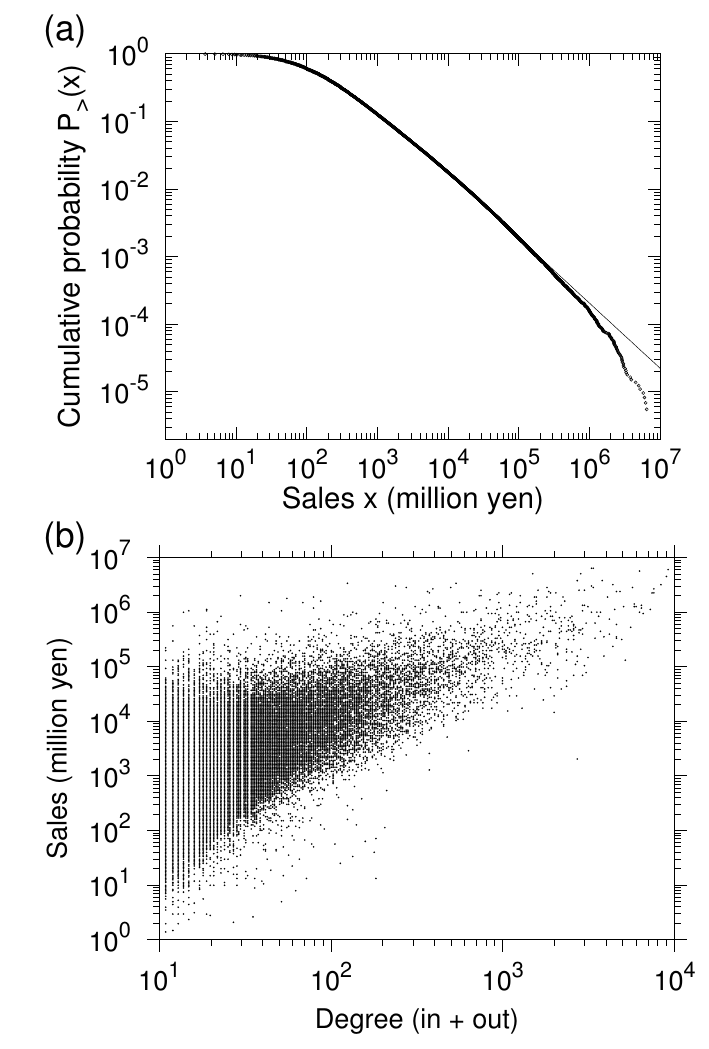}
  \caption{%
    (a)~Cumulative distribution for firm-size measured by sales.
     A power-law distribution $P_>(x)\propto x^{-\alpha}$ is fitted
     by MLE with $\alpha=0.96\pm0.02$ (a solid line).
    (b)~Scatter-plot for degree (total) and sales.}
  \label{fig:firm_size}
\end{figure}

The firm-size, however measured, obeys a power-law, being well known
as a Zipf's law. For the nodes in the network, we examined financial
data (availability exceeds 70\% presumably missing only extremely
small firms). The cumulative distribution for the sales of those nodes
(0.73 million) is shown in \figref{fig:firm_size}~(a). The Zipf's law,
$P_>(x)\propto x^{-\alpha}$, is obvious for sales $x$. The exponent is
close to unity, $\alpha=0.96\pm0.02$ by MLE estimated for $x>10^4$
million yen.

\figref{fig:firm_size}~(b) depicts a scatter-plot for total degree and
sales. Correlation between the firm's degree and size is
positive. The statistical significance can be quantified by non-parametric
statistics, such as Kendall's rank correlation, $\tau$
(see \cite{press1992nrc}). For the data, $\tau=0.391$ ($p$-value
$<10^{-7}$), which shows a significant positive correlation between the degree
and firm-size. We used different quantities for firm-size, such as
profits and the number of employees, and obtained very similar results. In
addition, when considering either of in- or out-degree, we can observe
that each has a positive correlation with firm-size.

\subsection{Transitivity}\label{sec:transitivity}

Unlike many social networks, the supplier of a firm's supplier is not
likely also to be the firm's supplier, and similarly for customer,
because such a process of production is redundant for most cases.
Transitivity means how high the number of triangles is present in the
network (see the review \cite{newman2003sfc}). Here we regard the
network as an undirected graph.

Global clustering coefficient is defined by
$C_{\text{g}}=(3\times\text{number of triangles})/%
(\text{number of connected triples})$,
where a connected triple means a pair of nodes that are connected
to another node. $C_{\text{g}}$ is the mean probability that two
firms who have a common supplier/customer are also suppliers/customers
of each other. The undirected graph of our dataset yields
\begin{equation}
  C_{\text{g}}=1.87\times10^{-3}=0.187\%\ .
  \label{Cg}
\end{equation}

To compare this value with that for a class of random graphs having a
same degree sequence but randomly rewired links, we use the expected
value of global clustering coefficient given by \cite{newman2003rgm}
\begin{equation}
  C_{\text{g}}=\frac{\avek}{N}\,\left[
    \frac{\avekn2-\avek}{\avek^2}\right]^2
  \ .
  \label{random_Cg}
\end{equation}
Putting the values \eref{k1}, \eref{k2}
and \eref{num_N} into \eref{random_Cg}, we have
$C_{\text{g}}=1.81\times10^{-2}$. The observed value \eref{Cg} is,
therefore, merely 10\% of \eref{random_Cg}, and shows weaker
transitivity than what is expected by chance. This is reasonable
because triangular relations, during the selection of suppliers and
customers, are suppressed in the formation of them.

The average of local clustering coefficient is, on the other hand,
equal to 4.58\% for the same dataset.

\subsection{Degree correlation}\label{sec:deg_corr}

For each node, the in-degree and out-degree are highly correlated.
This is consistent with what we saw in
Section~\ref{sec:firm_size} that each quantity has positive
correlation with firm-size.

For each link, to see the assortative mixing with respect to degrees
$(k_1,k_2)$ at both end of each link \cite{newman2003mpn}, or degree
correlation, let us examine the joint distribution for $(k_1,k_2)$.
Here we ignore the direction of links, but even when taking the possible
four combinations of in/out at a directed link, we obtain similar
results. To test for the assortativity, we calculate the frequency
$F(k_1,k_2)$ that the pair of $k_1$ and $k_2$ appears at either end of
a link in the network. Then compare it with a same quantity
$F_{\text{r}}(k_1,k_2)$ that is obtained in a randomized network with
the same degree sequence. We generated 1,000 randomized networks, and
quantify as the ratio $F/F_{\text{r}}$ where $F_{\text{r}}$ is the
average for the randomizations.

\begin{figure}[htbp]
  \includegraphics[width=0.34\textwidth]{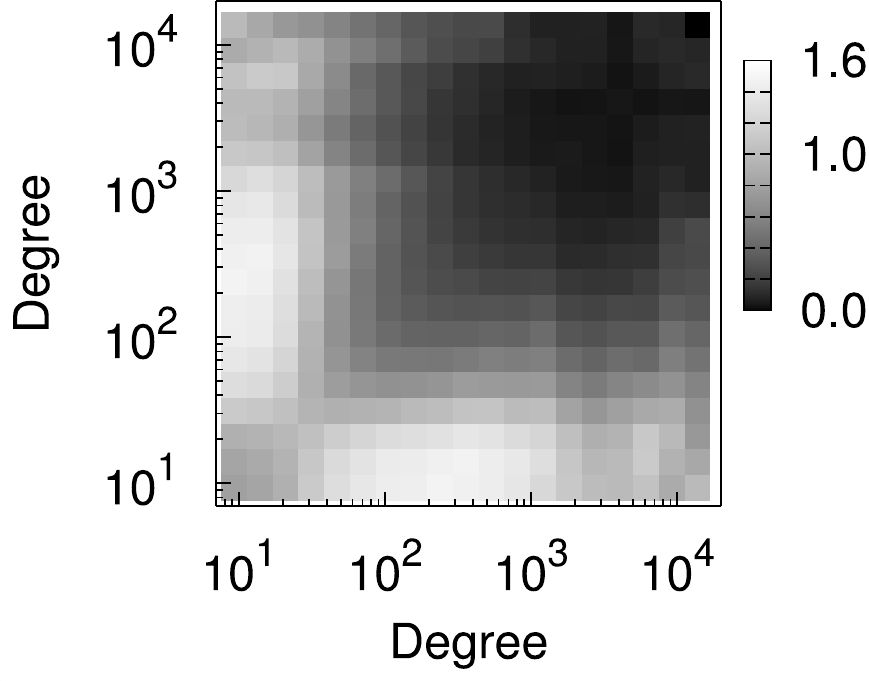}
  \caption{%
    Joint distribution for degrees (total) at end of each link. 
    The value is the ratio of the actual frequency divided by what is
    expected by chance in random networks with the same degree
    sequence as the actual one.}
  \label{fig:deg_corr}
\end{figure}

The result is shown in \figref{fig:deg_corr}. One can observe that
large-degree nodes, large firms, are connected with small-degree
nodes, small firms. For the hubs referred to in the end of
Section~\ref{sec:degree}, they have a large number of suppliers and
customers, but similarly for firms with intermediate-size, displaying
disassortativity \cite{newman2003mpn}. This can be quantified by the
Pearson correlation coefficient $r$ for $(k_1,k_2)$. For the data, we
have
\begin{equation}
  r=-0.0747\pm0.0002\ ,
\label{eq:pearson}
\end{equation}
where the error calculated by the method given in
\cite{newman2003mpn}. This claims that $r$ is negative with a
statistical significance.

\section{Community structure}\label{sec:commun}

The global connectivity examined in Section~\ref{sec:connectivity}
shows that basically all industries are highly entangled with each
other within the weakly or strongly connected component. Yet the
connectivity alone does not tell how dense or sparse the stream of
production is distributed depending on industrial or geographical
groups. Detection of community structure is to find how nodes cluster
into tightly-knit groups with high density in intra-groups and with
lower connectivity in inter-groups.

\begin{table*}
  \caption{\label{tab:community}%
    Communities extracted for the subgraph composed of manufacturing
    firms as nodes (about 0.12 million). Modularity optimization was
    recursively done for largest communities to obtain the
    sub-communities, ten of which are shown here. In each of them are shown ten
    firms with largest degrees are listed with names, major groups
    (primary/secondary/tertiary, if any in this order) of industrial
    sectors (see \tabref{tab:sect_major}), and sub-community sizes.}
\begin{ruledtabular}
\begin{tabular}{cp{2cm}p{15cm}}
  no. & annotation &
  firms (major groups; primary/secondary/tertiary), $\ldots$\hfill[community-size] \\
  \hline
  01 & heavy industry &
  Mitsubishi Heavy Industries (30/26),
  Kawasaki Heavy Industries (26/30),
  Kobe Steel (23/25),
  Ishikawajima-harima Heavy Industries (30/26),
  Sumitomo Heavy Industries (26),
  Nippon Steel (23),
  Kubota Industries (30/27/23),
  Mitsui Engineering and Shipbuilding (30),
  Hitachi Zosen Shipbuilding (26),
  Sumitomo Metal Industries (23),
  $\ldots$\hfill[7,447] \\
  02 & foods &
  Itoham Foods (09),
  Prima Meat Packers (09),
  Yamazaki Baking (09),
  Nisshin Seifun Flour (09),
  Maruha Nichiro Foods (09),
  Nippon Flour Mills (09),
  Q.P. Foods (09),
  Nihon Shokken Foods (09),
  Toyo Suisan Foods (09),
  Ichiban-foods (09),
  $\ldots$\hfill[7,115] \\
  03 & transportation equipment &
  Honda (30/27),
  Nissan (30),
  Toyota Motor (30),
  Aisin (25/30/27),
  Mitsubishi Motors (30),
  Denso (30/27),
  Fuji Heavy Industries (30),
  Toyota Industries (30/26),
  Suzuki Motor (30),
  Isuzu Motors (30),
  $\ldots$\hfill[5,769] \\
  04 & construction material &
  Sumitomo Osaka Cement (22),
  Air-Water Industrial Gas (17/18),
  Kyowa Concrete (22),
  Hokukon Concrete (22),
  Marukin Steel Materials (23),
  Mitsubishi Construction Materials (25/22),
  Hinode Steel/Manhole (23/22),
  Nihon Kogyo Industrial (22/13),
  Lafarge Aso Cement (22),
  Maeta Concrete (22),
  $\ldots$\hfill[2,644] \\
  05 & pulp/paper &
  Oji Paper (15),
  Rengo Paper (15),
  Nippon Paper (15),
  Oji Chiyoda Container (15),
  Tomoku Container (15),
  Morishigyo Paper (15),
  Settsu Carton (15),
  Morishigyo Paper Sales (15),
  Crown Package (15),
  Yamato-shiki Paper (15/19),
  $\ldots$\hfill[3,697] \\
  06 & electronics(a) &
  Hitachi (28/29/27),
  Fujitsu (32/28),
  NEC (28/29),
  TDK (27/29),
  Oki Electric (28/29),
  Hitachi High-Technologies (31/26),
  Rohm Semi-conductors (29),
  Murata Electronics (27),
  IBM Japan (28),
  Japan Radio Communication Equipment (28/27),
  $\ldots$\hfill[3,082] \\
  07 & electronics(b) &
  Matsushita (Panasonic) (27/31),
  Sharp (29/27/28),
  Sanyo (27/25),
  Panasonic Shikoku Electronics (29/27/28),
  Pioneer (27/28),
  Matsushita Battery (27),
  Sanyo Tottori (28),
  Matsushita Refrigeration (27/26),
  Kenwood (28),
  CMK Electronic Devices (29),
  $\ldots$\hfill[2,921] \\
  08 & electronics(c) &
  Canon (28/26/31),
  Seiko Epson (28/29),
  Omron (27),
  Nikon (31/26),
  Ricoh (26/28),
  Fujinon Optics (31),
  Hoya Optics (31),
  Casio (26/31/28),
  Pentax Optics (31/28),
  Sony EMCS Electronic (27/28),
  $\ldots$\hfill[2,692] \\
  09 & electronics(d) &
  Toshiba (27/28/29),
  Stanley Electric (27/26),
  Toshiba Lighting and Technology (27/26/29),
  Ushio Electric (25/27/26),
  Hamamatsu Photonics (29/27),
  Nippon Electric Glass (22),
  Toshiba Tec (26/27),
  GS Yuasa Industry (27/29),
  Iwasaki Electric (27),
  Topcon Electric (31),
  $\ldots$\hfill[2,320] \\
  10 & apparel &
  Renoun Apparel (12),
  Onward Kashiyama Apparel (12),
  MC Knit Apparel (12),
  World Apparel (12),
  Sanyo Shokai Apparel (12),
  Itokin Apparel (12),
  Fujii Fabrics (11),
  Sanei-International Apparel (12),
  YKK Fastening and Accesaries (32),
  World Apparel (12),
  $\ldots$\hfill[1,567] \\
\end{tabular}
\end{ruledtabular}
\end{table*}

\begin{figure*}
  \includegraphics[width=0.85\textwidth]{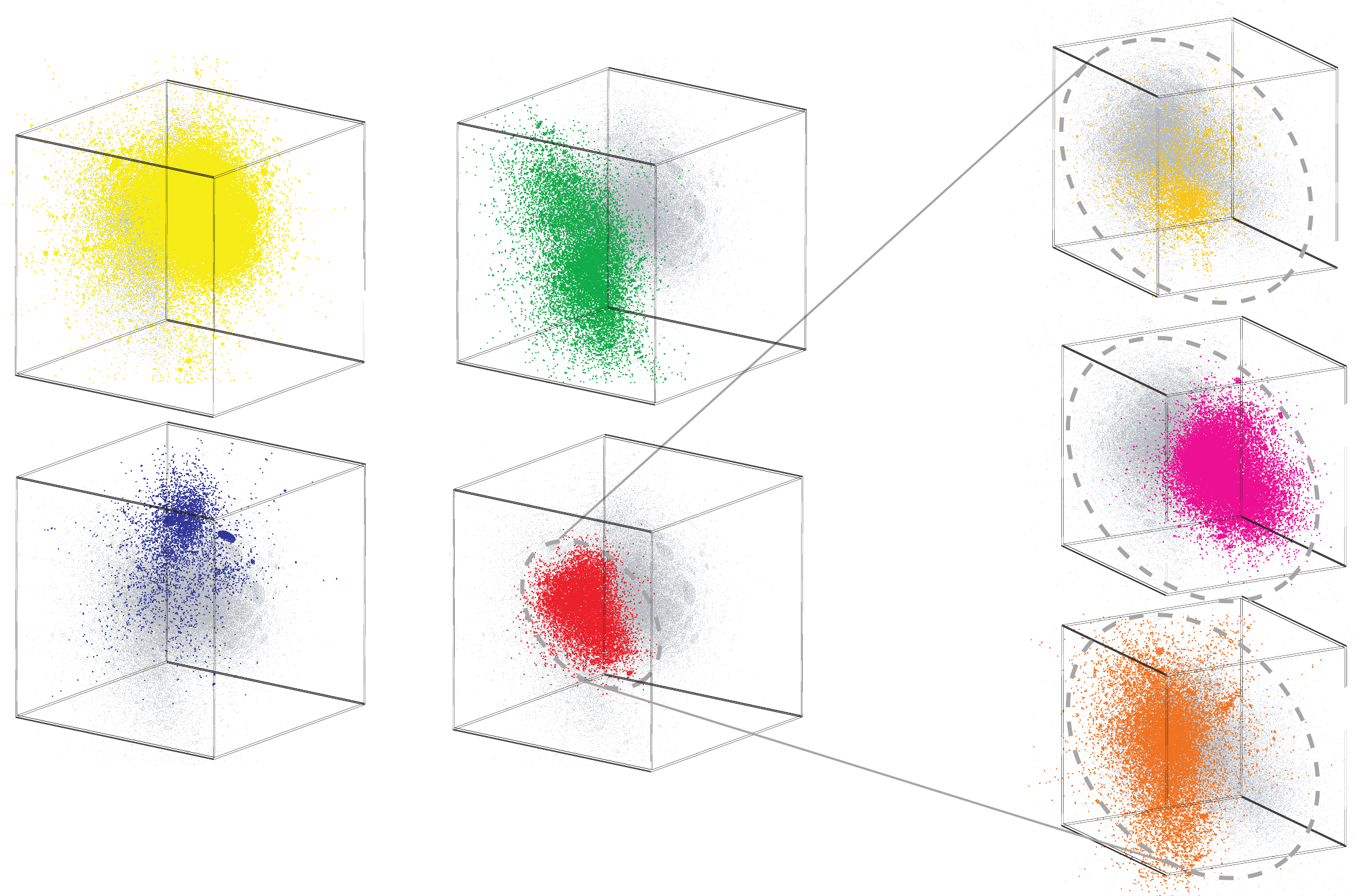}
  \caption{%
    (Color online) Layout of nodes for firms in the manufacturing sector (F) by a
    force-directed method. The links are omitted, and different colors
    are put on the nodes belonging to four largest communities.
    The community of color red (middle bottom and encircled by a dotted
    line) is divided into three sub-communities of electronics (a),
    (b) and (c) given in \tabref{tab:community} (enlarged in the right
    column).}
  \label{fig:lgl_pnet}
\end{figure*}

We focus, in this section, on the manufacturing sector with 0.12
million firms, in order to understand the sector's modular structure
by excluding other dominant sectors including wholesale and retail
trade, which obviously have a different role in the stream of
production from the core of manufacturing sector.

We use the method of maximizing modularity, introduced by
\cite{newman2004fae} and implemented for large-scale graphs in
\cite{clauset2004fcs} as a greedy optimization. While considerable
studies have been conducted to develop various methods for community
extraction, we use the modularity optimization for its clear
interpretation in terms of statistical hypothesis (also see
\cite{danon2005ccs} for a comparative study). Let $e_{ij}$ be the
fraction of edges in the network that connect nodes in group $i$ to
those in group $j$, and let $a_i\equiv\sum_j e_{ij}$, $b_j\equiv\sum_i
e_{ij}$. Then modularity $Q$ is defined by
\begin{equation}
  Q=\sum_i(e_{ii}-a_i b_i)
  \label{def_Q}
\end{equation}
which is the fraction of edges that fall within groups, minus the
expected value of the fraction under the hypothesis that edges fall
randomly irrespectively of the community structure. The method is
formulated as an optimization problem to find a partition of nodes
into mutually disjoint groups such that the corresponding value of $Q$
is maximum.

As shown in \cite{fortunato2007rlc,kumpula2007lrc}, however, the
method can give undesired grouping, depending on the density of
connections and the network size. Especially, large communities can
potentially contain sub-communities. Currently, without an established
method to avoid this problem of resolution limit (see
\cite{kumpula2007lrm}, for example, and also
\cite{lancichinetti2009doh}), we shall check the
structure of detected communities by constraining modularity
optimization on each single community, especially for those with
relatively large community-size.

We apply the method of community extraction to the undirected subgraph
whose nodes consist of only firms in the manufacturing sector
(division F in \tabref{tab:sect}). The resulting modularity
\eref{def_Q} is $Q=0.566\pm0.001$, which indicates
strong community structure (the error calculated by the method given
in \cite{newman2003mpn}). The number of extracted communities exceeds a
thousand, whose sizes range from a few to more than 10,000.  From the
database of the information on the firms, we found
that many of those small communities are each located in same geographical areas
forming specialized production flows. An example is a small group
of flour-maker, noodle-foods producers, bakeries, and packing/labeling
companies in a rural area.

On the other hand, five large communities exceed 10,000 each in size,
being possibly subject to the above problem of resolution. After checking the
sub-communities in the above mentioned fashion, we obtained the
communities as tabulated in \tabref{tab:community}. The necessity of
this procedure can be clearly seen for the communities of so-annotated
``electronics'' (a)--(d), which constitute a single community in the
first stage of optimization. Each firm is classified into one or more
industrial sectors, and the major-group classifications (2 digits; see
\tabref{tab:sect_major}). Obviously a community contains
those firms in closely related industrial sectors. The annotations ---
heavy industries, foods, transportation equipment, etc. --- are made
by such observations.

Let us closely examine the modular structure of those large communities.
Note that ten firms with the largest degrees (typically largest
firm-sizes) are listed in each community. We note that these large
firms in a same community do not form a set of nodes that are mutually
linked in nearly all possible ways, or a quasi-clique. Rather, with their suppliers and
customers, they form a quasi-clique in a corresponding bipartite
graph as follows. A supplier-customer link $u\rightarrow v$ for a set of nodes
$V$ ($u,v\in V$) can be considered as an edge in a bipartite graph
that has exactly two copies of $V$ as $V_1$ and $V_2$ ($u\in V_1$ and
$v\in V_2$). Those large and competing firms quite often share a set
of suppliers to some extent, depending on the industrial sectors,
geographical locations and so on.

For example, Honda ($v_1$), Nissan ($v_2$) and Toyota ($v_3$) possibly
have a number of suppliers $u_i$ of mechanical parts, electronic
devices, chassis and assembling machines, etc., in common. Then the
links form a clique or a quasi-clique in the bipartite graph, where
most possible links from $u_i$ to $v_1$, $v_2$, $v_3$, $\ldots$ are
present. This forms a portion in the original graph with a higher
density than other portions. By enumerating cliques in the bipartite
graph and examining them, we found that this is actually the case for
the community (03) in \tabref{tab:community}, and similarly for all the
other communities therein.

For the case of electronics (a)--(d), those quasi-cliques are further
separated into groups. Namely, the suppliers belong to different
groups of industrial organization for historical reasons and the
so-called {\it keiretsu\/}, and/or are located in divided geographical
sectors. The sub-communities (a)--(d) can be considered as such
separate groups with mutually sparse links. The electronics (b), for
instance, are originated and developed in an urban area in western
Japan, not in eastern urban area of the Tokyo, being different from
group (a).

These interpretations of modular structure should be strengthened by more
detailed analysis, especially with a new technique for extraction of
communities that are present in multi-scale levels in the hierarchical
organization of the production of network (see
\cite{salespardo2007eho,clauset2008hsa} for example), which is to be
published elsewhere.

Here, to check the intra-group and inter-group connectivities, we
resort to visualization of the entire manufacturing
sector by a graph layout based on a physical simulation. The system in
the simulation consists of point-particles for nodes and springs for
links. The springs obey Hooke's law with a spring constant, and the
particles have a Coulomb charge with a same sign, exerting
repulsive forces inversely proportional to the square of mutual
distances, for nodes to spread well on the layout. A resistance force
is also acting on each particle, being proportional to its velocity,
in order to relax the system in a final layout.
The Barnes-Hut tree algorithm \cite{barnes1986hnl} is
employed for fast computation, and the Coulomb interaction was
calculated on a special-purpose device (GRAPE; gravity
pipeline) invented for astrophysical $N$-body simulation
\cite{makino1998sss}. The result is depicted in \figref{fig:lgl_pnet}.
Details of the layout method is given in \cite{fujiwara2009vls}.

One can observe that nodes within a tightly-knit group cluster at
mutually near positions in the layout, while different communities are
separated from each other with overlapping portions. The sub-communities
stated above appear as clusters nested in the community of
electronics. Also even closer look in enlargement (not shown in the
figure) shows blobs corresponding to hubs or large firms associated
with their suppliers and customers.

\section{Chain of bankruptcy}\label{sec:bankrupt}

Let us now turn our attention to the dynamics on the production network.
Firms put value added on intermediate goods in {\it anticipation\/}
of gaining profits --- anticipation, because no firm knows how much
their produced goods might actually be demanded by other firms and
consumers. In addition, they face uncertainty in the change of
costs for intermediate goods to purchase as inputs, as well as in
fluctuations of labor and financial costs. Only {\it a posteriori\/},
therefore, a firm's profit, being equal to sales minus costs, is
determined through the interaction with others in the network.

Supplier-customer link is a credit relation \cite{stiglitz2003tnp}.
Whenever one delivers goods to others without an immediate exchange
of money or goods of full value, credit is extended. Frequently,
suppliers provide credit to their customers, who supply credit to
their customers and so forth. Also customers can provide credit to
their suppliers so as to have them produce an abundance of intermediate goods
beforehand. In either case, once a firm goes into financial
insolvency state, its creditors will possibly lose the scheduled
payment, or goods to be delivered that have been necessary for
production. The influence propagates from the bankrupted customer to
its upstream in the former cases, and similarly from the bankrupted
supplier to its downstream in the latter cases. Thus a creditor has
its balance-sheet deteriorated in accumulation, and may eventually go
into bankruptcy. This is an example of a {\it chain of bankruptcy\/}.

A bankruptcy chain does not occur only along the supplier-customer
links. Ownership relation among firms is another typical possibility
for such creditor-debtor relationship. It is, however, also frequently
observed in our dataset that supplier-customer links are also present
between holding and held companies, and siblings and affiliated firms.
We assume that most relevant paths along which the chain of bankruptcy
occurs are the creditor-debtor links of the production network.

As explained in Section~\ref{sec:data}, we have an exhaustive list of
bankruptcies. Corresponding to the snapshot of the network taken in
September 2006, we employ all the bankruptcies for exactly one-year
period from October. The number of bankruptcies amounts to roughly
0.013 million, daily mean being 30, and includes a few bankruptcies of
listed firms. Nearly half of the bankrupted firms, precisely
$N_{\text{b}}\equiv 6264$, were present on the network at the
beginning and went into bankruptcy during the period.  The rest are of
extremely small-size, typically with one employee, and were not
included as nodes, which we assume irrelevant to our purpose as well
as new entry of firms during the same period.

Let us define the probability of bankruptcy in the one-year period by
the ratio between the number of bankrupted nodes, $N_{\text{b}}$, and
the initially present nodes, $N$ given by \eref{num_N}, by
\begin{equation}
  p=N_{\text{b}}/N\approx 0.620\%\ .
\label{eq:a}
\end{equation}
Note that the probability has inverse of time in its physical
dimension. A year was chosen for the time-scale so that it should be
longer than the time-scale for financial activities of firms,
typically weeks and months, and be shorter than that for the change of
network itself.

\subsection{Avalanche-size distribution}

Let us first take a look at how a certain size of chain of
bankruptcies actually takes place. Here a chain is defined as a set of
bankrupted nodes that are connected by links that are present in the
initial network. If nodes are white and black according to
survival and bankruptcy during the period, a chain
means connected black nodes surrounded by white nodes,
and its size refers to the number of black nodes in the chain.
Note that a chain is not necessarily a path in the graph, but can be a
tree, and may include cycles in it.

\begin{figure}[htbp]
  \includegraphics[width=0.46\textwidth]{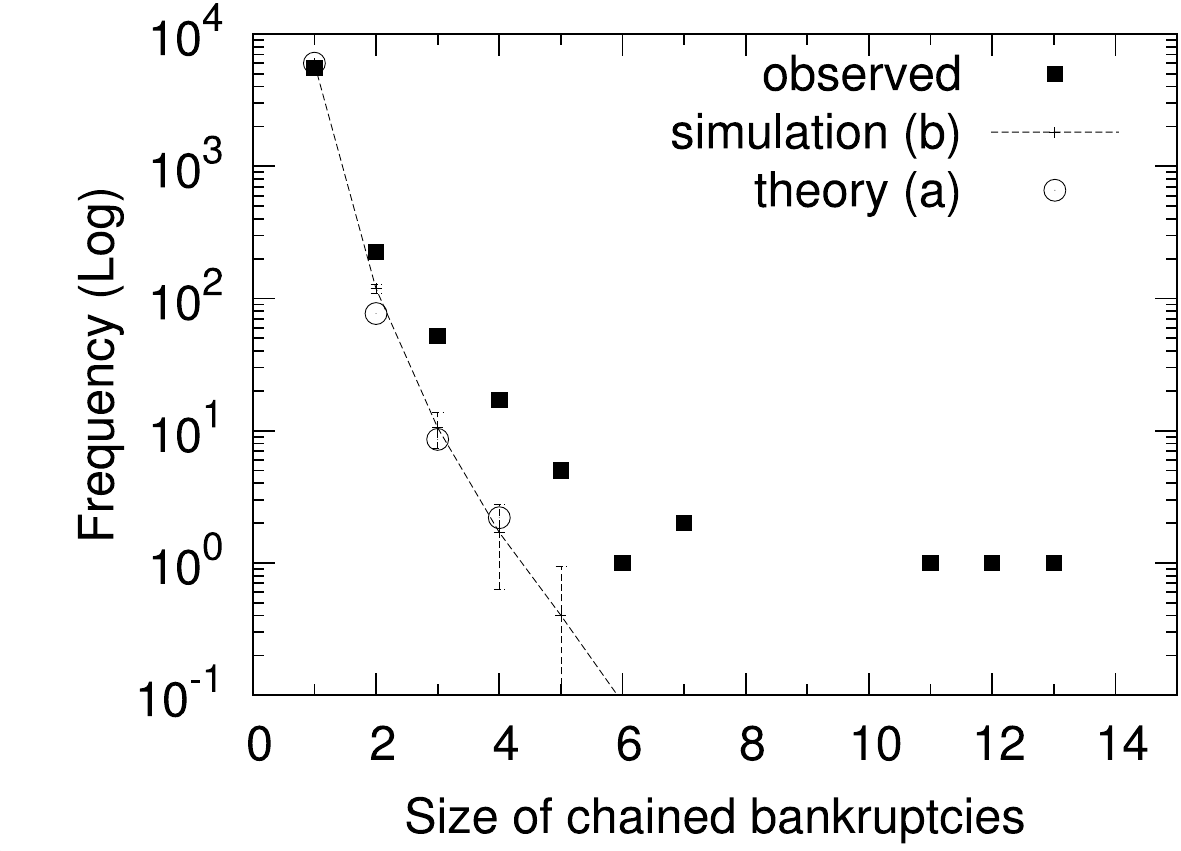}
  \caption{%
    Frequency (vertical in log-scale) of avalanches with a specific
    size (horizontal in linear). Filled squares are the observed
    frequencies in the observation. Open circles show a
    theoretical calculation for randomized networks with anonymous
    nodes (see (a) in the text), and a line with error bars represents a
    Monte-Carlo calculation for randomized networks with same
    bankrupted nodes as observed ones (b).}
  \label{fig:avalan_size}
\end{figure}

\figref{fig:avalan_size} shows the size-distribution of such
avalanches by filled squares, which represents the frequency
distribution of avalanches with a specific size. The observed values
are tabulated in \tabref{tab:akky}.

\begin{table}
  \caption{\label{tab:akky}%
    Comparison between the actual value of the
    chain-bankruptcy and the values expected from coincidence.
    ``Obs." for the observed values,
    ``Theor." for the theoretical values of coincidence,
    ``O/T" for the ratios between them,
    while ``RNW" is the value obtained by simulation for
    the randomized network.}
  \begin{ruledtabular}
    \begin{tabular}{crrcl}
      & \mbox{Obs.} & \mbox{Theor.} & \mbox{O/T} &
      \mbox{RNW}\footnotemark[1] \\
      \hline
      $\quad B_1\quad$ & 5507 & 6013 & 0.9 & 5985(21) \\
      $B_2$ & 226 & 76.9 (82.3\footnotemark[2]) & 2.9 & 118.5(10.0) \\
      $B_3$ & 52 & 8.6\footnotemark[2] & 6.0 & 10.5(3.2) \\
      $B_4$ & 17 & 2.2\footnotemark[2] & 7.7 & 1.7(1.1)
    \end{tabular}
  \end{ruledtabular}
  \footnotetext[1]{Standard deviations in parentheses (each 1,000 randomizations).}
  \footnotetext[2]{Mean-field approximations.}
\end{table}

\subsection{Evaluation of accidental chain}

Let us then evaluate how a certain size of chain of actual
bankruptcies occurs more or less frequently than what is expected
simply by chance. Suppose, in a random network with a specified degree
sequence, one selects $Np$ nodes for failure, where $p$ is the
probability of failure per node. Then calculate the frequency of a
certain number of failed nodes that are connected by links, and we can
compare the frequency of accidental chain of failures with that for an
actual chain of bankruptcies. We shall use the terms, {\it failures\/}
and {\it failed nodes\/}, to mean a set of black nodes that are
selected randomly and uniformly over all the nodes, and
to distinguish them from the set of actually observed
bankruptcies and bankrupted nodes.

The selection of failed nodes can be done in two ways, that is, (a) by
choosing uniformly random nodes to fail irrelevantly from the actual
data of nodes, or alternatively (b) by specifying exactly the same
bankrupted nodes in the actual data, but in otherwise randomized
network with the same degree sequence as the real one. These two ways
possibly yield different results for our purpose, so let us perform
the evaluation in both ways. The evaluation (a) allows us to
understand how the accidental chain is related to the network
properties, especially degree distribution and correlation, the
results of which were given in Section~\ref{sec:stat}. On the other
hand, we can take into account of difference between failures and
bankruptcies in our terminology here. We elaborate the calculation of
(a) in the following Section~\ref{sec:eval}. For the calculation of
(b), the estimates are obtained by a Monte Carlo simulation generating
random networks with failures associated to the actually bankrupted
nodes.

We denote by $B_m$ the number of clusters, each with $m$ failed nodes
that are connected by links, in average for randomized networks with
the same degree distribution $P(k)$ as the actual production network.
Since the clustering coefficient is of the same order of magnitude as
what is expected by chance, we assume that the network is tree-like.
We shall calculate $B_m$ for $m=1,2,3$ and 4 in order.

The results are summarized in \tabref{tab:akky} and
\figref{fig:avalan_size} for the actually observed values of $B_m$
along with the evaluated values based on the above-mentioned two
classes of randomized networks, (a) and (b), in the columns ``Theor.''
and ``RNW'' respectively. We find that (a) and (b) give quantitatively
similar estimates.  By comparing the actually observed values with the
evaluation for random networks with a same degree sequence, we can
conclude that the avalanche size has a much heavier tail in its
distribution for size larger than 3. Those large avalanches involve
regionally and industrially related firms, as we could confirm from
our dataset.  Therefore, the vulnerable paths, along which a chain of
bankruptcy takes place are present in those modular groups. The
following Section~\ref{sec:eval} is devoted to the explanation of the
evaluation (a).

\subsection{Evaluation for accidental chain of failures}\label{sec:eval}

We count the number of accidental chain of failures in a given,
tree-like network, where all the failures occur randomly with
probability $p$ per node.

\subsubsection{$\bm{B_1}$: Isolated failure}

\begin{figure}[htbp]
  \centering
  \includegraphics[width=0.38\textwidth]{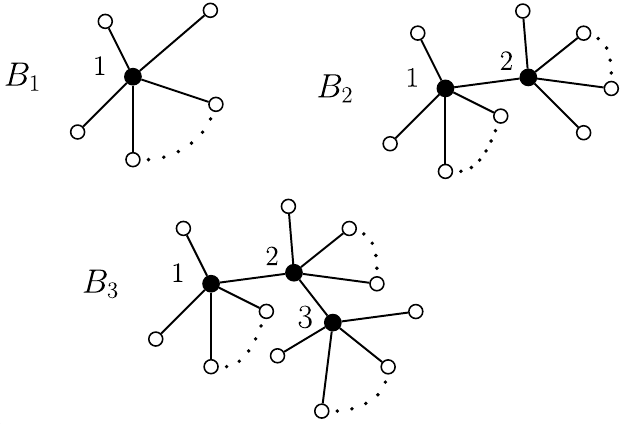}
  \caption{%
    Clusters of $m$ failed nodes that are connected by links, which
    contribute to $B_{1,2,3}$. Black nodes are failed
    ones, and white nodes are non-failed. The numbers attached with
    failed nodes correspond to the subscripts of degrees, $j$ for
    $k_j$.}
  \label{fig:b1}
\end{figure}

Denoting the number of nodes of degree $k$ by $K_1(k)$, it is related
to the degree distribution $P(k)$ by
\begin{equation}
  P(k)=\frac1N K_1(k)\ .
\end{equation}
Obviously
\begin{equation}
  \sum_{k=1}^\infty K_1(k)=N\ .
\end{equation}
Among $K_1(k)$ such nodes, the average number of failed nodes is $p
K_1(k)$. Since the probability that all the nodes connected to a
failed node are not in failure is $(1-p)^k$ (see \figref{fig:b1}), the
average number of isolated failure is given by
\begin{align}
  B_1 &= \sum_{k=1}^\infty p\,K_1(k)\,(1-p)^{k}=p\,NR_1\ ,
  \label{B1}\\
  R_1&\equiv \ave{(1-p)^{k}}\ ,
  \label{R1}
\end{align}
where $\ave{\cdot}$ means average over the nodes, {\it i.e.},
\begin{equation}
  \ave{f(k)}\equiv
  \frac{\displaystyle\sumone f(k) K_1(k)}{\displaystyle\sumone K_1(k)}\ .
  \label{avedef1}
\end{equation}
Note that since $Np$ is equal to the actual number of the
bankruptcies, $R_1$ gives the rate of the isolated failure. From the
observed distribution of $K_1(k)$ and $p\approx 0.006204$, we obtain
\begin{gather}
  R_1\approx 0.9600\ ,\label{r1res}\\
  B_1\approx 6013.2\ .
  \label{b1res}
\end{gather}
The actual number of the {\it isolated\/} bankruptcies is 5,507, being
92\% of this estimate. Following the standard argument for
$1/\sqrt{n}$ estimate of the statistical errors, we see that there are
less number of isolated bankruptcies in actuality than that expected
by chance.

The results \eref{r1res} and \eref{b1res} apply to any class of
networks with the same degree sequence $K_1(k)$, in particular,
irrespectively of degree correlation.

\subsubsection{$\bm{B_2}$}

We denote by $K_2(k_1,k_2)$, the number of pairs of nodes; each pair
of nodes having degree $k_1$ and $k_2$. Precisely, choose a node with
degree $k_1$, and count the nodes with degree $k_2$ connected with the
first node. After doing this over all the nodes and by adding up the
resulting numbers, one has the quantity $K_2(k_1,k_2)$. Now the number
of double-failure case, $B_2$, can be expressed by $K_2(k_1,k_2)$ as
follows:
\begin{align}
  B_2 &= \frac12\sumtwo K_2(k_1,k_2) \, p(1-p)^{k_1-1}\,p(1-p)^{k_2-1}
  \nonumber\\
  &= \frac12 \frac{p^2}{(1-p)^2} \,R_2 \sumtwo K_2(k_1,k_2)\ ,
  \label{b2res}\\
  R_2&\equiv \aves{2}{(1-p)^{k_1+k_2}}\ ,
  \label{r2def}
\end{align}
where the combinatorial factor $1/2$ accounts for the over-counting
the chain in the reverse order of $k_1$ and $k_2$, and $\aves{2}{\cdot}$
denotes the average over the links, defined by
\begin{equation}
  \aves{2}{f(k_1,k_2)}\equiv\frac{\displaystyle\sumtwo 
    K_2(k_1,k_2) f(k_1,k_2)}{\displaystyle\sumtwo K_2(k_1,k_2)}\ .
  \label{avedef2}
\end{equation}

From the definition, $K_2(k_1,k_2)$ satisfies the identities:
\begin{gather}
  K_2(k_1,k_2) =K_2(k_2,k_1)\ ,
  \label{k101}\\
  \sum_{k_2=1}^\infty K_2(k_1,k_2) =k_1 K_1(k_1)\ .
  \label{k1sumsum}
\end{gather}
The two identities lead to the summation formula:
\begin{equation}
  \sumtwo K_2(k_1,k_2) =N\avek\ ,
  \label{k2mo1}
\end{equation}
which is exactly twice the number of links, as it should be.
The following identity is also useful.
\begin{equation}
  \sumtwo K_2(k_1,k_2) k_2^n =N\avek \ave{k_2^n}_2=N\avekn{n+1}\ .
  \label{k2formula}
\end{equation}
Using Eq.~\eref{k2mo1}, $B_2$ can be put as
\begin{equation}
  B_2 = \frac12 \frac{p^2}{(1-p)^2}\,R_2 N\avek\ .
  \label{B2res1}
\end{equation}

The factor $R_2$ can be calculated directly from the actual values of
$K_2(k_1,k_2)$. The result is
\begin{equation}
  R_2\approx 0.488\ ,
  \label{R2exact}
\end{equation}
which leads to
\begin{equation}
  B_2 \approx 76.9\ .
  \label{b2res2}
\end{equation}
The observed value is about three times of this estimate, as shown in
\tabref{tab:akky}, which indicates the double-failure chain is much more
abundant than what is expected by chance.

\rann

In the case of random network, the estimation reduces to a simple
expression. In fact, first note that $K_2(k_1,k_2)$ can be written in
terms of $K_1(k)$ as follows:
\begin{equation}
  K_2^{\text{(ran)}}(k_1,k_2)= \frac1{N\avek} K_1(k_1)\,k_1 k_2\,K_1(k_2)\ ,
  \label{k2mf}
\end{equation}
because it is equal to the number of nodes of degree $k_1$, $K_1(k)$,
multiplied by the probability of choosing the node of degree $k_2$,
$k_2 K_1(k_2)/\sum_{k_2=1}^\infty k_2 K_1(k_2)$.
Note that \eref{k2mf} satisfies the identities \eref{k101} and \eref{k1sumsum}
as is required for the consistency of the calculation. 
Inserting \eref{k2mf} into \eref{r2def} and \eref{avedef2}, the value of $R_2$ in
this approximation is given by
\begin{equation}
  R_2^{\text{(ran)}}= R_{11}^2\ ,
  \label{R2ran}
\end{equation}
where 
\begin{gather}
  R_{11}\equiv \frac{\ave{k (1-p)^k}}{\avek}\approx 0.723\ .
  \label{r11val}
\end{gather}
This yields
\begin{equation}
  R_2^{\text{(ran)}} \approx 0.523\ ,
  \label{R2ranv}
\end{equation}
and
\begin{equation}
  B_2^{\text{(ran)}}=\frac12 \,\frac{p^2}{(1-p)^2}\, R_2^{\text{(ran)}} N \avek \approx
  82.3\ ,
\end{equation}
as being tabulated in \tabref{tab:akky}.

\subsubsection{$\bm{B_3}$}

We define $K_3(k_1,k_2,k_3)$ in a similar way for $K_2(k_1,k_2)$; take
a node with degree $k_1$, continue the counting to nodes with degree
$k_2$ and then $k_3$ (see \figref{fig:b1}). $B_3$ is given by
\begin{align}
  B_3&=\frac12 \sumthree K_3(k_1,k_2,k_3) \nonumber\\
  &\times p(1-p)^{k_1-1} p(1-p)^{k_2-2} p(1-p)^{k_3-1}
  \label{B3ori}
  \nonumber\\
  &= \frac12 \frac{p^3}{(1-p)^4}\,R_3\,K_3^{\text{(sum)}} \ ,\\
  R_3&\equiv \aves{3}{(1-p)^{k_1+k_2+k_3}}\ ,\\
  K_3^{\text{(sum)}}&\equiv\sumthree K_3(k_1,k_2,k_3)\ ,
\end{align}
where the combinatorial factor $1/2$ is to cancel the over-counting,
and $\aves{3}{\cdot}$ refers to the average weighted with
$K_3(k_1,k_2,k_3)$ in the same manner as that in Eq.~\eref{avedef1}
and \eref{avedef2}.

By definition,
$K_3(k_1,k_2,k_3)$ satisfies the following identities:
\begin{gather}
  K_3(k_1,0,k_3) =0\ ,\label{k3i1}\\
  K_3(k_1,k_2,k_3) = K_3(k_3, k_2, k_1)\ ,\label{k3i2}\\
  \sum_{k_3=1}^\infty K_3(k_1,k_2,k_3) =
  K_2(k_1,k_2)(k_2-1)\ .\label{k3i3}
\end{gather}
Using the identities \eref{k3i3} and \eref{k1sumsum}
we find that
\begin{equation}
  K_3^{\text{(sum)}}
  =N\left(\avekn2-\avek\right)
  \approx 3.09\times10^9 .
  \label{k3sum}
\end{equation}

\rann

Since exact evaluation of $R_3$ involves the evaluation of $K_3(k_1,k_2,k_3)$,
which requires a huge computational resource, let us evaluate $R_3$ by
using a random-network approximation.
By considering attaching the nodes \#2 and \#3 successively
to the node \#1 with equal probability on each links as in the case of
$K_2^{\text{(ran)}}$, we obtain
the following:
\begin{align}
&  K_3^{\text{(ran)}}(k_1,k_2,k_3) = \nonumber\\
&  \frac1{N^2\avek^2} K_1(k_1) k_1 k_2 K_1(k_2) (k_2-1) k_3K_1(k_3)\ ,
  \label{k3ran}
\end{align}
which satisfies identities \eref{k3i1}--\eref{k3sum},
except that $K_2$ is replaced by $K_2^{\text{(ran)}}$ in \eref{k3i3}.
By using the above in \eref{B3ori} we obtain;
\begin{equation}
R_3^{\text{(ran)}}=
R_{11}^2
\frac{R_{12}\avekn2 -R_{11}\avek}{\avekn2 -\avek}
\end{equation}
where 
\begin{equation}
  R_{12}\equiv \frac{\ave{k^2 (1-p)^k}}{\avekn2}\approx 0.0450\ .
\end{equation}
This leads to
\begin{align}
  R_3^{\text{(ran)}} &\approx 0.0226\ ,\\
  B_3^{\text{(ran)}} &\approx 8.55\ .
\end{align}

\subsubsection{$\bm{B_4}$}

\begin{figure}[htbp]
  \centering
  \includegraphics[width=0.37\textwidth]{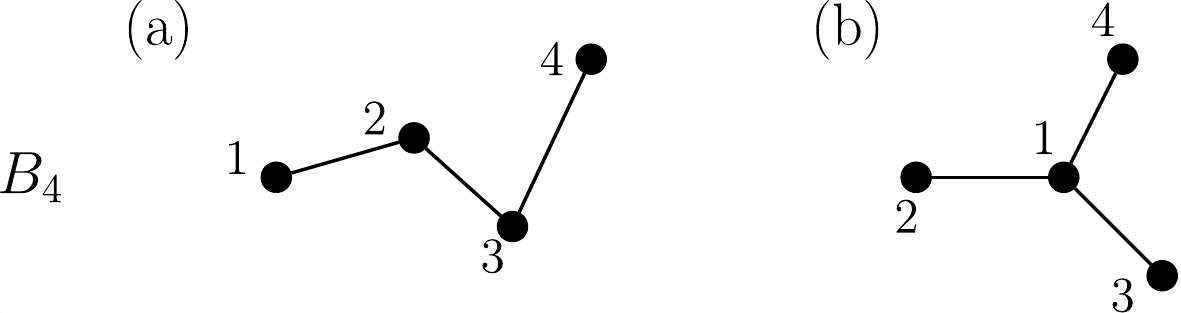}
  \caption{%
    Two types of clusters that contribute to $B_4$. Non-failed
    nodes (white nodes in \figref{fig:b1}) are not drawn.}
  \label{fig:b4}
\end{figure}

The clusters that contribute to $B_4$ are illustrated in
\figref{fig:b4}, and are divided into two types as depicted.
One has to understand that larger clusters are more rare events so
that statistical errors in observation increase drastically.
With this in mind, let us perform estimation, and compare them with
the observed values.

\subsubsection*{Type (a)}

Contribution of the cluster (a) can be written as follows using the
number of strings $k_1$--$k_4$, $K_4(k_1,k_2,k_3,k_4)$;
\begin{align}
  B_{4a}&=\frac12\,  p^4 \sumfour K_4(k_1,k_2,k_3,k_4) 
  \nonumber\\
  &\times (1-p)^{k_1-1}\biggl(\,\prod_{j=2}^3(1-p)^{k_j-2}\biggr)(1-p)^{k_4-1}
  \nonumber \\
  &=\frac12\frac{p^4}{(1-p)^6}R_{4a}\,K_4^{\text{(sum)}}\ ,  \label{B4}\\
R_{4a}&\equiv \langle(1-p)^{\sum_{i=1}^4 k_i}\rangle_4 \ ,
\end{align}
where definitions of $\langle \cdot \rangle_4$ and $K_4^{\text{(sum)}}$
should be self-evident.

Let us first calculate $K_4^{\text{(sum)}}$. The identities satisfied 
by $K_4(k_1,k_2,k_3,k_4)$ are similar to those 
of $K_3(k_1,k_2,k_3)$ and would be now obvious. Using them,
the summations over $k_1$ and $k_4$ can be carried out as follows,
\begin{equation}
  K_4^{\text{(sum)}}
  =\sum_{k_2,k_3=1}^\infty K_2(k_2,k_3) (k_2-1)(k_3-1)
  \label{k4sum}
\end{equation}
In the expansion of each summand, all the terms except for those of
$k_2 k_3$ allow further summation by repeatedly using the identities
given so far. On the other hand, since the coefficient of the $k_2
k_3$-term is $K_2(k_2,k_3)$, this term should be related to
degree correlation. As usual, we define the correlation coefficient
$r_1$ by
\begin{equation}
  r_1\equiv
  \frac{\ave{k_1 k_2}_2 - \ave{k_1}_2^2}{\ave{k_1^2}_2-\ave{k_1}_2^2},
  \label{rdef}
\end{equation}
Using this definition, we can write that
\begin{align}
  \ave{k_1 k_2}_2
  &=r_1\avekn2_2+(1-r_1){\avek_2^2}
  \nonumber\\
  &=r_1\frac{\avekn3}{\avek}+(1-r_1)\left(\frac{\avekn2}{\avek}\right)^2\ ,
  \label{r1k1k2}
\end{align}
where we used Eq.~\eref{k2formula} and re-labeled the subscripts in
the degrees. Thus the $k_2 k_3$-term reads
\begin{align}
  \sumtwo& k_1 k_2 K_2(k_1,k_2)
  = N\avek \ave{k_1 k_2}_2
  \nonumber\\
  =&N
  \left( r_1\avekn3+(1-r_1)\frac{\avekn2^2}{\avek}\right).
  \label{k4sum2}
\end{align}
Putting all the terms together, we have
\begin{equation}
  K_4^{\text{(sum)}}
  = N
  \left( r_1\avekn3+(1-r_1)\frac{\avekn2^2}{\avek}
    -2\avekn2+\avek \right).
  \label{k4sum3}
\end{equation}

The random-network approximation for this case requires a
careful treatment because of the appearance of the degree correlation coefficient
$r_1$ in the above summation formula.
Although its value given in \eref{eq:pearson} is small, 
it has a critical role in the above equations:
If we use the random-network approximation for $K_2$ given
in \eref{k2mf}, we obtain the $r_1=0$ result;
\begin{equation}
  \sumtwo k_1 k_2 K_2^{\text{(ran)}}(k_1,k_2)
  = N \frac{\avekn2^2}{\avek}\approx 1.188\times10^{12},
\label{toolarge}
\end{equation}
while the exact value is
\begin{equation}
  \sumtwo k_1 k_2 K_2(k_1,k_2)
  \approx 1.342\times10^{11}.
\label{k2exact}
\end{equation}
The role of the correlation coefficient $r_1$ is evident in these values;
it brings in partial cancellations between the first term and
the second term, so that the actual value is much smaller than
that of the random-network value \eref{toolarge}.
Note that this is deeply connected with the asymptotic behavior of
the degree distribution noted in Section~\ref{sec:degree}:
If all the moments of degree is of order one, the effect of the
correlation coefficient $r_1$ is not this drastic.
However, due to the degree distribution being power-law,
the moments $\avekn2$ and $\avekn3$ are proportional to a positive power of $N$
(we will elaborate on the analysis in Appendix~ref{sec:appA})
and thus are quite large, resulting in the importance of
cancellation by $r_1$ observed above.  

For this reason, we evaluate $B_{4a}$ in two schemes in the following.
The first scheme is to use the exact distribution for $K_2(k_2,k_3)$ but
use the random-network approximation for $k_1$ and $k_4$, so that
\eref{k2exact} is satisfied. The second one is to use the
random-network approximation to all the nodes in $K_4$.
We will carry out the calculation of both schemes separately.

\rannn1

The first approximation scheme is given by
\begin{align}
  & K_4^{\text{(ran1)}}(k_1,k_2,k_3,k_4)=\frac1{N^2\avek^2}
  \nonumber\\
  & \times K_1(k_1) k_1(k_2-1) K_2(k_2,k_3) (k_3-1) k_4 K_1(k_4)\ ,
  \label{k4ran1}
\end{align}
which is obtained by attaching the \#1 and \#4 nodes to a \#2--\#3
pair randomly. It is evident that this satisfies the identity \eref{k4sum} and
therefore \eref{k4sum3}. It then follows that 
\begin{align}
&  R_{4a}^{\text{(ran1)}}=
  \frac{R_{11}^2}{K_4^{\text{(sum)}}} \nonumber\\
& \times
  \sum_{k_2,k_3=1}^\infty (1-p)^{k_2+k_3}(k_2-1)(k_3-1)
  K_2(k_2,k_3)\nonumber \\
& \approx 8.33\times10^{-3} \ .
\end{align}
From this we obtain
\begin{equation}
  B_{4a}^{\text{(ran1)}}\approx 0.819\ .
\end{equation}

\rannn2
In the second, complete random-network approximation, we have
\begin{align}
  & K_4^{\text{(ran2)}}(k_1,k_2,k_3,k_4) =\frac1{N^3\avek^3} 
\nonumber \\
&\times K_1(k_1) k_1 k_2 K_1(k_2) (k_2-1) k_3 K_1(k_3) (k_3-1) k_4 K_1(k_4)\ ,
  \label{k4ran}
\end{align}
which is obtained by connecting the node \#2,3,4 in sequence,
or alternatively, by substituting the random network approximation $K_2^{\text{(ran)}}$ in
\eref{k2mf} for $K_2$ in \eref{k4ran1}.
We then obtain the following:
\begin{align}
R_{4a}^{\text{(ran2)}}&=R_{11}^2
\left(\frac{R_{12}\avekn2-R_{11}\avek}{\avekn2-\avek}\right)^2\nonumber\\
&\approx 9.77\times10^{-4} \ ,
\end{align}
which leads to
\begin{equation}
  B_{4a}^{\text{(ran2)}}   \approx 0.884\ ,
\end{equation}
which is very close to $B_{4a}^{\text{(ran1)}}$.

\subsubsection*{Type (b)}

For the other type of (b), we can write as
\begin{align}
  B_{4b} &= \frac1{3!}\, p^4 \sumfour J_4(k_1,k_2,k_3,k_4) 
  \nonumber\\
  &\times
  (1-p)^{k_1-3}\prod_{j=2}^4 (1-p)^{k_j-1}
  \nonumber\\
  &\equiv \frac1{3!}\frac{p^4}{(1-p)^7}\,R_{4b}\,J_4^{\text{(sum)}}\ ,
\label{B4b}
\end{align}
where 
\begin{equation}
  J_4^{\text{(sum)}}\equiv \sumfour J_4(k_1,k_2,k_3,k_4)\ ,
\end{equation}
and $R_{4b}$ is a ratio defined by the above.
In this case, we denote by $J_4(k_1,k_2,k_3,k_4)$, the number of the
clusters of type (b) with the degrees $k_i$ of nodes \#$i$ in
\figref{fig:b4}. The combinatorial factor $1/3!$ cancels the
over-counting of a same cluster. The following identities hold:
\begin{gather}
  J_4(k_1,k_2,k_3,k_4)=
  J_4(k_1,k_{\sigma(2)},k_{\sigma(3)},k_{\sigma(4)}),
\label{ji1}
  \\
  \sum_{k_4=1}^\infty J_4(k_1,k_2,k_3,k_4)
  =K_3(k_2,k_1,k_3)(k_1-2),
\end{gather}
where $\sigma(j)$ represents a permutation of $j=2,3,4$. Using the
identities, we have
\begin{align}
  \sum_{k_3,k_4=1}^\infty J_4(k_1,k_2,k_3,k_4)=K_2(k_2,k_1)(k_1-1)(k_1-2)\ ,
  \label{jsum1}
\end{align}
which leads to
\begin{align}
  J_4^{\text{(sum)}}=
  N\left(\avekn3-3\avekn2+2\avek\right)\ .
  \label{jsum2}
\end{align}

\rann

As seen in \eref{jsum2}, the degree-correlation does not play any 
major role for this type of cluster. So, unlike the case of $B_{4a}$, 
let us employ a simple random-network approximation of the form:
\begin{align}
  &J_4^{\text{(ran)}}(k_1,k_2,k_3,k_4)=\frac{1}{N^3\avek^3}K_1(k_1) 
  \nonumber\\
  &\times
  k_1 k_2\,K_1(k_2)
  (k_1-1) k_3\,K_1(k_3)
  (k_1-2) k_4\,K_1(k_4)\ ,
\end{align}
which satisfies identities \eref{ji1}--\eref{jsum2}
with $K_2$ replaced by $K_2^{\text{(ran)}}$.
We obtain the following:
\begin{align}
  R_{4b}^{\text{(ran)}}&=R_{11}^3
  \frac{R_{13}\avekn3-3R_{12}\avekn2+R_{11}\avek}{\avekn3-3\avekn2+\avek}
  \nonumber\\
  &\approx 3.52\times10^{-4},
\end{align}
where 
\begin{equation}
  R_{13}\equiv \frac{\ave{k^3 (1-p)^k}}{\avekn3}
  \approx 9.57\times10^{-4}
\ .
\end{equation}
This leads to
\begin{equation}
  B_{4b}^{\text{(ran)}}\approx 1.39\ .
\end{equation}

\newpage
\section{Discussion}\label{sec:discussion}

In the preceding section, we found that the size of chained
bankruptcies is frequently larger than what is expected simply by
chance. This fact implies that supplier-customer links can be
potentially a vulnerable path for a chain of bankruptcies causing a
sequence of failures, and that it is important to do monitoring and
prediction of such failures on a nation-wide scale. Let us remark a
few things concerning the finding.

Firstly, the study on community structure in Section~\ref{sec:commun}
showed that the firms are often tightly knitted into clusters
characterized by closeness in industrial sectors and/or geographical
locations. Because those firms in such a cluster are susceptible to
common risk factors specific to the sector or the location, the
bankruptcies in a cluster can be correlated due to the similar
profiles of those firms, but not necessarily due to the ``link''
effect.

It is not easy, in the present analysis, to separate the link effect
from the sectoral or locational correlation, but yet we can roughly
estimate the relative importance of the link effect. Although it is
difficult to determine a single cause in each bankruptcy, there exists
a pattern of types in the cause of failure, which are categorized and
compiled in the database. The most frequent one is the poor
performance in business, typically slow sales and decreasing profit
due to adverse and sluggish market conditions. The other main category
is the linked failure due to a secondary effect from bankruptcy of
customer, subsidiary or affiliated companies, which often causes the
loss of accounts receivable. See \cite{fujiwara2008cfb} for the
details.

For the bankrupted nodes $N_{\text{b}}$ studied in
Section~\ref{sec:bankrupt}, the categorized causes of bankruptcy are
the poor performance (60\%) and the link effect (7\%).  But if a pair
of bankrupted nodes are connected by a supplier-customer link, the
link effect is considerable (27\%), while the poor performance cases
or ``solo'' failures are relatively less frequent (42\%). This fact,
therefore, shows that the chain of bankruptcy we observed in the
preceding section is largely, even if not entirely, due to the link
effect taking place along the supplier-customer links.

Secondly, the chain of bankruptcy has a great influence in a
nation-wide economy. In fact, the total amount of debts for bankrupted
firms in a year typically ranges from 10 to 25 trillion yen in the
last 10 years, roughly equal to more than 100 billion euro. This
amounts to 2\% or even more of the nominal GDP in Japan. Of course,
all the debts are not to be lost, but it should not be undervalued the
fact that there are a large number of creditors who have given credits
to those bankrupted firms. Therefore, the study on the chain of
failures and its ripple effect has practical applications. Recent
models such as a model of credit chains and bankruptcies
\cite{battiston2007ccb}, simulation of avalanche effect
\cite{lubloy2008network}, Potts-like model of contagion
\cite{sieczka2009cfb} (see also references in \cite{schweitzer2009en})
would provide valuable insights and tools to do monitoring of
financial fragility and prediction of such failures on a nation-wide
scale.

Thirdly, from a broader perspective, the production network has
a similarity in its structure with other economic networks. While the
inter-bank networks have unique structure among financial institutions
\cite{demasi2006fmi,iori2008nai,kyriakopoulos2009nea}, and the
banks-firms networks is basically a bipartite network
\cite{demasi2009ajc}, the ownership networks
\cite{garlaschelli2005sew,glattfelder2009bcn} possess similar
properties of network. In particular, the community structure has
industry-sectoral modules in a hierarchical way. However, the common
sets of supplies play important constituents in the production
network, as we have explained by the example of automobile companies
in Section~\ref{sec:commun}, and this is completely different from the
ownership networks. We would yet need a more systematic comparison
with other economic networks for the study of similarity and
dissimilarity.

\section{Summary}\label{sec:summary}

We studied a large-scale structure of the nation-wide production
network comprising a million firms and four million
supplier-customer links in Japan. The set of nodes covers most active
firms. Each link was chosen and considered as important, in a
systematic survey of credit informations, by at least one of the firms
at either end of the link, as its suppliers and customers. We found
scale-free degree distribution, disassortativity, correlation of
degree to firm-size, and small clustering coefficients compared with
randomized networks with the same degree sequence. In the community
analysis, which is based on modularity optimization, we were able to
identify communities in the manufacturing sector, and found that they
can be interpreted as modules depending on industrial sectors and
geographical regions. Large communities contained subgroups that can
be characterized also by industrial organization and development.

In addition, by employing an exhaustive list of bankruptcies that took
place on the production network, we took a close look at the size
distribution for chains of bankruptcies, or avalanche-size
distribution. We elaborated a method to evaluate the frequencies of
accidental chain in randomized networks, and found that the actual
avalanche has a heavy tail distribution in its size. Combining with
the large-scale properties and heterogeneity in modular structures, we
claim that the effect to a number of creditors, non-trivially large due
to the heavy tail in the degree distribution, is considerable in the
real economy of the nation.

\begin{acknowledgments}
  We would like to thank Tokyo Shoko Research, Ltd.~in Japan for
  kindly providing us with a chance to study their large datasets for the
  academic purpose. Network analysis was partially carried out on
  Altix3700 at YITP in Kyoto University. Y.~F. thanks Y.~I.~Leon
  Suematsu for an implementation of fast community extraction,
  Y.~Fujita and A.~Kawai for a large-graph layout by GRAPE (gravity pipe).
  We would like to thank Y.~Ikeda, H.~Iyetomi, J.~Kert{\'e}sz, M.~Gallegati, W.~Souma for
  useful comments and additional references.
\end{acknowledgments}

\appendix

\section{Analytic estimates of $R$'s and the asymptotic behavior of the degree distribution}
\label{sec:appA}

Since the probability of failure $p$ is small, one might want to
utilize a perturbative evaluation of the ratios, $R$'s.
Indeed such an analytical expression would be helpful in 
understanding what essentially determines the rate of the 
chain-bankruptcy.
In this Appendix, we show that the asymptotic behavior of the 
degree distribution plays the key role. 

Let us denote the probability density function (pdf) of
the degree $k$ by $P(k)$
and its cumulative distribution function (cdf) by
\begin{equation}
P_>(k)=\sum_{k'=k}^\infty P(k')\ .
\end{equation}
We parametrize the cdf as
\begin{equation}
  P_>(k)\sim \left(\frac{k}{k_0}\right)^{-\mu}\ ,
  \label{pcum}
\end{equation}
for large $k$. It follows that $P(k)\propto k^{-\mu-1}$ in the same
region. For our data of the production network regarded as an
undirected graph, we have
\begin{gather}
  \mu\approx 1.366\ , \label{valuemu}\\
  k_0 \approx 2.18\ ,
  \label{valuek0}
\end{gather}
as maximum likelihood estimate (with the standard error of $\mu$ being
0.099). 

We define the generating function for the degree distribution by
\begin{equation}
  G(q)=\sum_{k=1}^\infty e^{-qk} P(k)\ ,
\end{equation}
which satisfy $G(0)=1$.
The desired ratios are expressed in terms of the generating function
$G(q)$ as
\begin{align}
R_1&=\frac{G(q_0)}{\avek}\ , \nonumber\\
R_{1n}&=-\frac{G^{(n)}(q_0)}{\avekn{n}}\ , \nonumber
\end{align}
where $G^{(n)}(q)$ is the $n$-th derivative of $G(q)$
with respect to $q$, and 
$q_0=-\log(1-p)\approx 6.20\times 10^{-3}$.

One might attempt the following analytic expansion of $G(q)$:
\begin{align}
  G(q)&=\sum_{k=1}^\infty \left(1- qk+\frac12 q^2k^2+\dots\right) P(k)
  \nonumber\\
  &=1-\avek q+\frac12\avekn2 q^2+\dots\ .
  \label{wrongp2}
\end{align}
This turns out to be {\it not} a useful expansion as is shown in what
follows. Instead, we shall give an improved expansion.

\begin{figure}[htbp]
  \centering
  \includegraphics[width=0.44\textwidth]{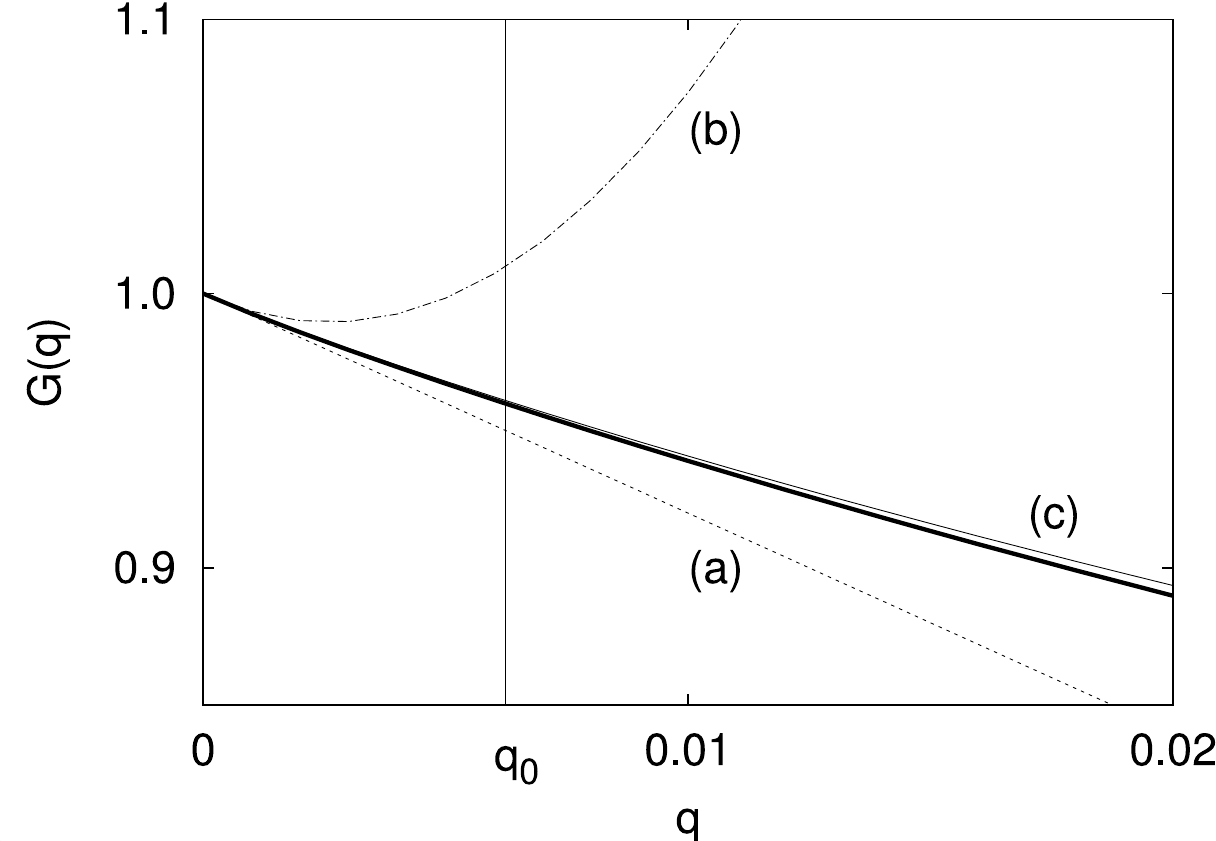}
  \caption{%
    The generating function for the degree distribution, $G(q)$ for
    $q\approx 0$. The solid curve is the actual plot. The first-order
    and the second-order approximations in Eq.~\eref{wrongp2} are
    shown by the dashed line~(a) and the dash-dot line~(b)
    respectively. The dotted line~(c) is the improved expansion given
    by \eref{improvede}. The vertical line corresponds to the
    actual value of $q_0=-\log(1-p)$.}
  \label{fig:p_expansion}
\end{figure}

For the distribution \eref{pcum}, the second moment of degree is
divergent for $\mu<2$ in a network with an infinite size. It is finite
but has a large value for network of a finite size. 
Actually, for our data 
\begin{equation}
  \avekn2=3069.6\ ,
\end{equation}
while $\avek=8.006$. So the expansion to the second order is a good
approximation only for
\begin{equation}
  q\ll\frac{\avek}{\avekn2}\approx 2.61\times 10^{-3}\ ,
\end{equation}
but this does not hold in the present case. This is illustrated in
Fig.~\ref{fig:p_expansion}, where the solid curve is the actual $G(q)$,
the curve (a) the first two terms in the expansion \eref{wrongp2},
the curve (b) all three terms in the expansion \eref{wrongp2}.
 
Let us now estimate the order of the coefficients of the naive expansion
\eref{wrongp2} analytically.
The $m$-th moment of degree is dominated by the large $k$ region
for $m>\mu$ as
\begin{align}
  \avekn{m} &\propto \sum^{\kmax} k^m k^{-\mu-1} \nonumber\\
  & \simeq \int^{\kmax} k^m k^{-\mu-1} dk 
  \simeq \kmax^{m-\mu}\ .
\end{align}
On the other hand, by considering the
node of the largest degree being the top of the
cdf \eref{pcum}, we have
\begin{equation}
  \kmax^{-\mu}\propto \frac1N\ .
\end{equation}
Therefore, we obtain
\begin{equation}
  \avekn{m}\propto N^{-1+\frac{m}{\mu}}\ ,
  \label{kmn}
\end{equation}
for $m>\mu$. It follows from \eref{kmn} that the $m$-th term in
\eref{wrongp2} is of order,
\begin{equation}
  \frac{N^{-1}}{m!}\left(N^{1/\mu}q\right)^m\ .
\end{equation}
Therefore the $m$-th order term is of the same order of magnitude as
the $(m+1)$-th order term provided that
\begin{equation}
  m\simeq N^{1/\mu} q\approx 154.7\ ,
\end{equation}
meaning that we need much more than 155
terms for the expansion \eref{wrongp2} to be useful for evaluation
of our ratios.

An improved approximation can be obtained as follows. Let us extract an
analytic contribution of the power-law tail by means of an analytic
continuation:
\begin{equation}
  \int^\infty \mu \frac{k^{-\mu-1}}{k_0^{-\mu}} e^{-qk} dk
  \simeq \mu (q k_0)^\mu \Gamma(-\mu)\ .
\label{gammaint}
\end{equation}
For $1<\mu<2$, this contribution is of larger power of $p$ than that
of the second-order, $p^2$, term in \eref{wrongp2}. Therefore, we
arrive at the following approximation,
\begin{equation}
  G(q)=1-\avek q + \mu\, \Gamma(-\mu) (k_0 q)^\mu +\dots\ .
  \label{improvede}
\end{equation}
Alternatively, this expression can be obtained by evaluating
the dominant $k\sim\kmax$ contribution in $G(q)-(1-\avek q)$.
Also it should be noted that this expression is valid for
 $q\gg 1/\kmax$, since we extended the integration to $k=\infty$
in \eref{gammaint}, instead of cutting it off at $k=\kmax$.
The curve (c) in Fig.~\ref{fig:p_expansion} depicts 
the behavior of the first three terms on the right-hand side of
\eref{improvede}.
It is evident that the improved
expansion works as an excellent approximation as shown in the plot.
In fact, the comparison between the estimates of the ratios obtained from \eref{improvede}
and the true values are excellent as seen in Table.~\ref{tab:goodgood}.

\begin{table}
\caption{The true values and the estimates obtained from
\eref{improvede}.}
\label{tab:goodgood}
\begin{tabular}{|c|r|r|r|}
\hline
Ratio & Exact value & Estimate & Difference\\
\hline
$R_1$ & 0.9600 & 0.9611 & 0.11\% \\
$R_{11}$ & 0.7230 & 0.7017 & -2.9\% \\
$R_{12}$ & $4.501\times 10^{-2}$ & $4.575\times 10^{-2}$ &1.6\% \\
$R_{13}$ & $9.574\times 10^{-4}$ & $9.440 \times 10^{-4}$ &-1.3\% \\
\hline
\end{tabular}
\end{table}

\bibliographystyle{unsrt}
\bibliography{lsspnet}

\end{document}